# In-Line-Test of Variability and Bit-Error-Rate of HfO$_x$-Based Resistive Memory

B. L. Ji[1], H. Li[1], Q. Ye[1], S. Gausepohl[1], S. Deora[2], D. Veksler[2], S. Vivekanand[1], H. Chong[1], H. Stamper[1],
T. Burroughs[1], C. Johnson[1], M. Smalley[1], S. Bennett[1], V. Kaushik[1], J. Piccirillo[1], M. Rodgers[1],
M. Passaro[1], and M. Liehr[1]
[1]College of Nanoscale Science and Engineering, SUNY Polytechnic Institute, Albany, NY 12203, USA
[2]SEMATECH, Albany, NY 12203, USA
E-mail: bji@sunycnse.com

*Abstract*—Spatial and temporal variability of HfOx-based resistive random access memory (RRAM) are investigated for manufacturing and product designs. Manufacturing variability is characterized at different levels including lots, wafers, and chips. Bit-error-rate (BER) is proposed as a holistic parameter for the write cycle resistance statistics. Using the electrical in-line-test cycle data, a method is developed to derive BERs as functions of the design margin, to provide guidance for technology evaluation and product design. The proposed BER calculation can also be used in the off-line bench test and build-in-self-test (BIST) for adaptive error correction and for the other types of random access memories.

## I. INTRODUCTION

In addition to performance and energy consumption, device variability has attained a critical importance over the past years for the technological evaluation of nano-electronic devices [1]. Both spatial (device to device) and temporal (cycle to cycle) variations are important for RRAM [2], memristor logic [3], and hybrid devices [4]. The write cycle variability of resistive memory is a particular challenge for technology robustness. To give early guidance to technologists and product designers, we propose a method to extract write BERs from electrical in-line-testing (ILT). Manufacturing variability on various levels (such as lot, wafer or chip) will also be discussed.

## II. MANUFACTURING VARIABILITY OF FORMING

Test structures of TiN/Ti/HfO$_x$/TiN or W/Ti/HfO$_x$/TiN RRAM stacks with various device sizes are fabricated at CNSE's 300 mm wafer fab. The RRAM device is crossbar-patterned with HfO$_x$ film (4~5nm thick) sandwiched between two electrodes. The bottom electrode (BE) is TiN. The top electrode (TE) consists of a TiN or Tungsten layer atop of a thin (3~6nm) Ti layer. Fig. 1 is a TEM cross section of a 50nm×50nm crossbar stack of TiN/Ti/HfO$_x$/TiN RRAM. Manufacturing variability of RRAM forming voltage (Vform) is characterized.

Using in-line-test results from 4 consecutive line monitor lots (Fig. 2), components of Vform variability for various devices are summarized in TABLE I. The overall coefficient of variation measured is about 6% for 50nm and 100nm devices.

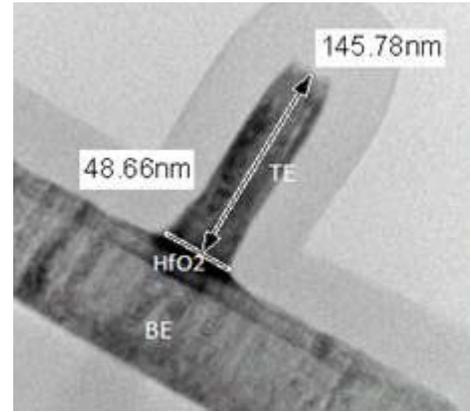

Fig. 1. TEM cross section of crossbar patterned RRAM stack. The designed size of cross bar is 50nm×50nm.

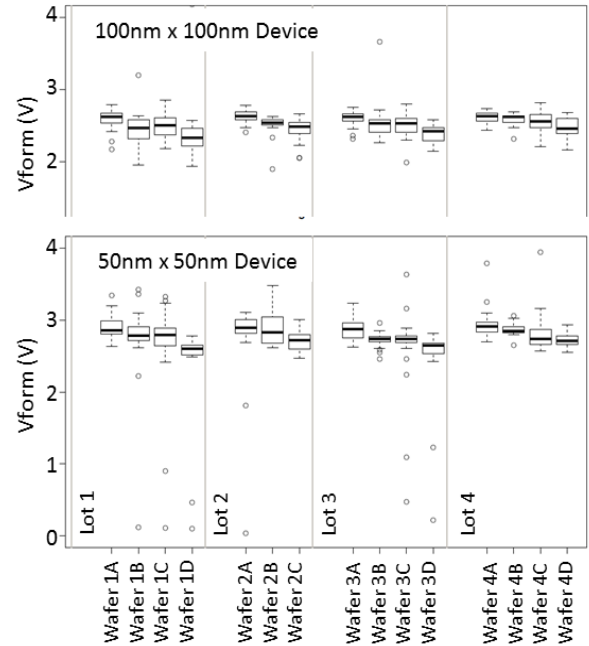

Fig. 2. Box plots of forming voltages of 50nm×50nm and 100nm×100nm RRAM from 4 consecutive lots.

TABLE I. FORMING VOLTAGE (VFORM) VARIABILITY AND ITS COMPONENTS FOR VARIOUS CROSSBAR DEVICES

| Parameter \ Device Size | 50nm×50nm | 100nm×100nm | 200nm×200nm |
|---|---|---|---|
| Vform | 2.78 V | 2.54 V | 2.18 V |
| $SD_{c2c}$ (chip-to-chip)[a] | 0.14 V | 0.13 V | 0.16 V |
| $SD_{w2w}$ (wafer-to-wafer) | 0.08 V | 0.04 V | 0.13 V |
| $SD_{l2l}$ (lot-to-lot) | 0.04 V | 0.04 V | 0.07 V |
| $SD_{total}$ | 0.17 V | 0.15 V | 0.22 V |
| $SD_{total}$ / Vform | 6% | 6% | 10% |

a. Standard deviation of the non-logarithmized values is denoted SD in this paper.

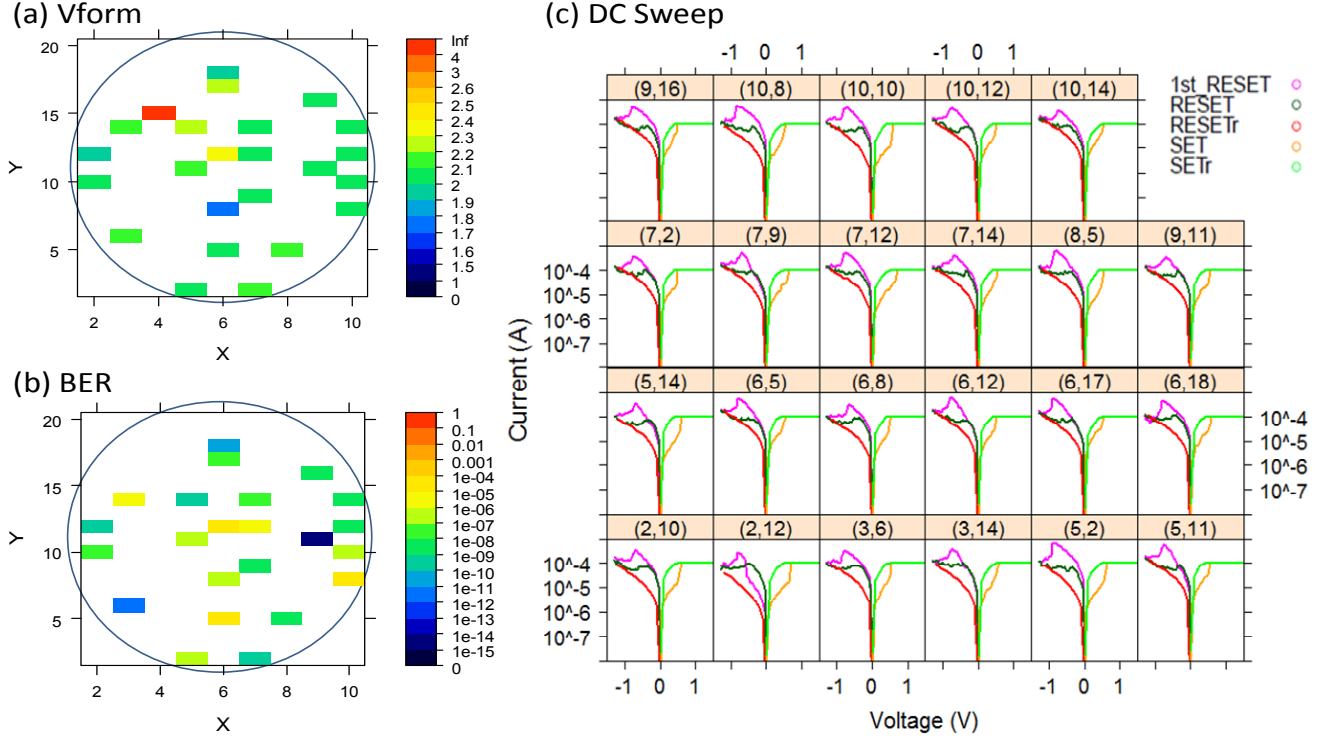

Fig. 3. Wafer maps of 100nm devices, (a) forming voltages, (b) BER. (c) After forming, DC sweeps are done in the sequence of first reset, set and reverse, reset and reverse. The chip coordinates are shown as (x,y) at the top banners of subpanels in Figs. 3c, 4 and 5.

The total variation is dominated by the chip-to-chip variation component, which also indicates reasonable stable process controls between wafers and lots.

### III. WRITE CYCLE VARIATION AND BIT ERROR RATE ANALYSIS

Nonvolatile memory system capacity is limited by the bit-error-rate [5]. Due to underlying physics, the temporal variation of a given RRAM technology may be difficult to reduce even as the manufacturing control advances. For system and circuit designers, it would be helpful to understand BER performance as early as possible.

Agilent 4073 parametric testers are used for all ILT operations including DC IV sweeps and pulsed voltage cycle tests. All pulses are set to 1 μs, which is the shortest holding time of the parametric testers. Fig. 3a is a tested wafer map of Vform of 100nm×100nm RRAM devices. RRAM forming is successful in 23 of 24 measured chips, but one chip (shown by red for Vform > 4V) is defective and removed from the following analysis. After forming, DC sweeps (Fig. 3c) and write operation on pulsed cycles (Fig. 4) are made. RRAM is switched to Low-Resistance-State (LRS) by a set voltage pulse (Vset = 1.8V) with a compliance current of 100 μA, or switched to High-Resistance-State (HRS) by a reset voltage pulse (Vreset = -1.5V). The resistance values in HRS and LRS are measured at 0.1V and will be denoted $R_H$ and $R_L$. The write cycle data are shown in both the temporal plot (Fig. 4) and the logarithmized Quantile-Quantile (Q-Q) plots (Fig. 5) for each chip. The Q-Q plots are fitted with straight lines to show strong log-normality.

Maximum likelihood fitting is used to fit ILT data to the respective log-normal distributions per chip,

$$R_H \sim lnN(\mu_H, \sigma_H^2) \quad (1)$$
$$R_L \sim lnN(\mu_L, \sigma_L^2) \quad (2)$$

where μ and σ are the location and scale parameters of log-normal distribution denoted as $lnN$. The 95% confidence interval of σ is 30% for 20 cycles. Chip-to-chip variations (of $\mu_L$, $\mu_H$, $\sigma_L$ and $\sigma_H$) within the tested wafer are shown in Fig. 6.

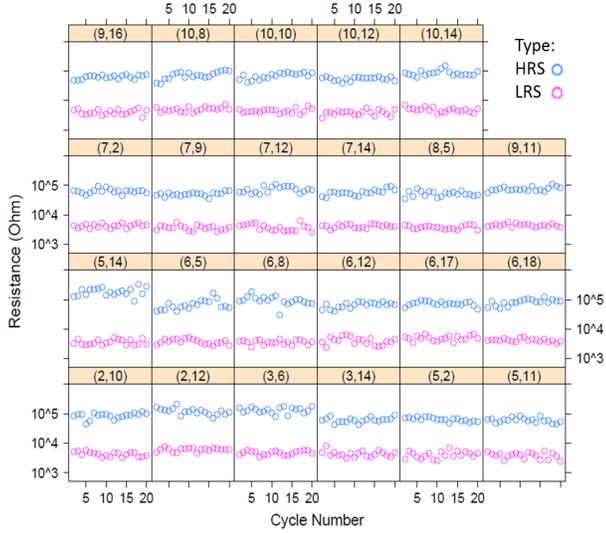

Fig. 4. HRS and LRS resistances measured in pulsed write cycles following forming and DC sweeps (Fig. 3c).

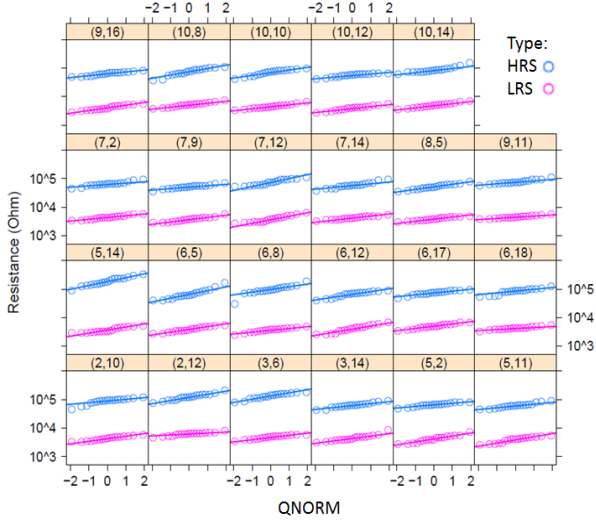

Fig. 5. Quantile-Quantile plots of the logarithmized resistance data (Fig. 4). Log-normal distribution is observed in each of the 23 measured chips.

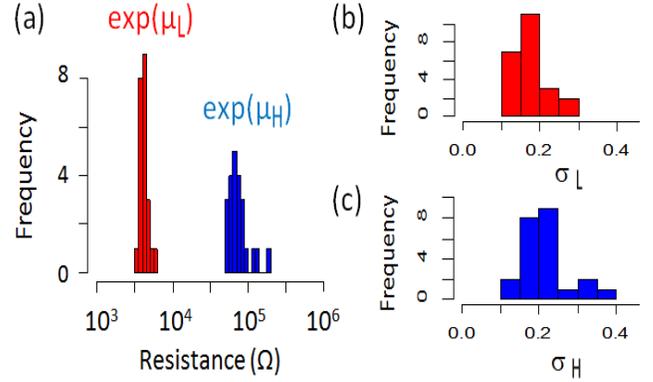

Fig. 6. Histograms of the log-normal fitting parameters for the measured HRS and LRS cycle distributions for 23 chips.

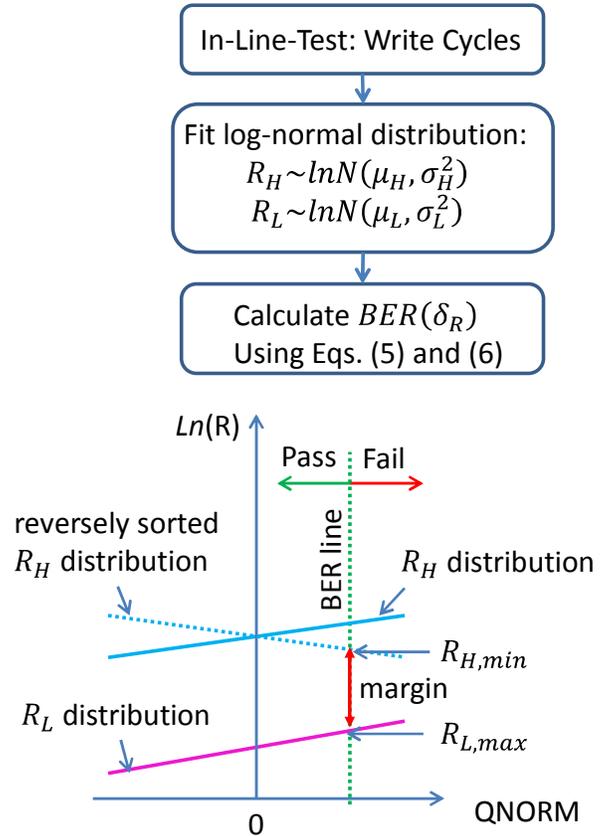

Fig. 7. Procedure from ILT to BER characterization. QNORM is the probability quantile function.

We propose a new approach for BER characterization at the device level (Fig. 7). BER curves are functions of the design margin $\delta_R$ that we define by the following equation,

$$\delta_R = (R_{H,min} - R_{L,max})/R_{L,max} \quad (3)$$

Used for sensing margin in the read circuit, (3) is similar to magnetic-RAM's MR ratio. The intrinsic BER is maximized when $R_{H,min}$ and $R_{L,max}$ are optimally chosen when the probabilities of failing both states are equal, so we have

$$BER(\delta_R) = P(R_H \leq R_{H,min}) = P(R_L \geq R_{L,max}) \quad (4)$$

In the case of log-normal distribution, (4) leads to,

$$ln(R_{L,max}) = \frac{\sigma_H \cdot \mu_L + \sigma_L \cdot \mu_H - \sigma_L \cdot ln(1+\delta_R)}{\sigma_H + \sigma_L} \quad (5)$$

$$BER(\delta_R) = \frac{1}{2}\left(1 - erf\left(\frac{ln(R_{L,max}) - \mu_L}{\sigma_L \sqrt{2}}\right)\right) \quad (6)$$

where *erf*(x) is the Gauss error function. Substituting (5) into (6), BERs are obtained as functions of the design margins. At the design margin of 100% ($\delta_R = 1$), the wafer map of the derived BER is shown in Fig. 3b, and the BER histogram and cumulative % for the tested wafer is shown in Fig. 8. As a function of design margins, BERs of the median, 25th and 75th percentile chips in the tested wafer are shown in Fig. 9. BER of about $10^{-8}$ is realized on the median chip at the design margin of 100% ($\delta_R = 1$).

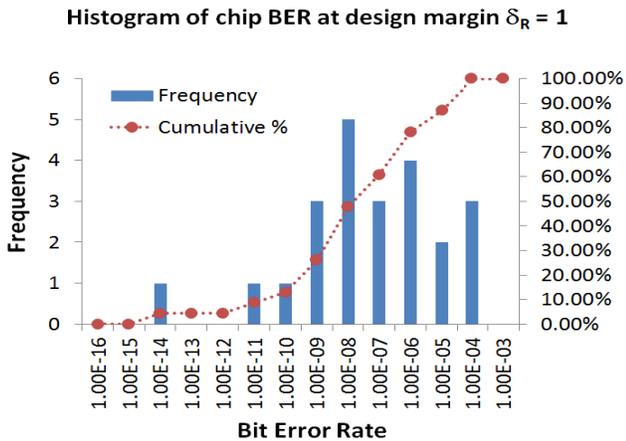

Fig. 8. Histogram and cumulative% of measured BERs at design margin of 100%.

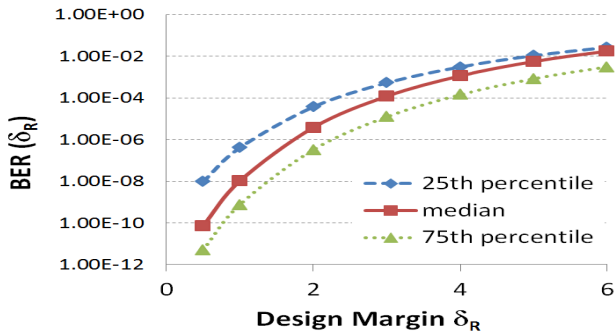

Fig. 9. BER curves of the median, 25th and 75th percentile chips in the tested wafer.

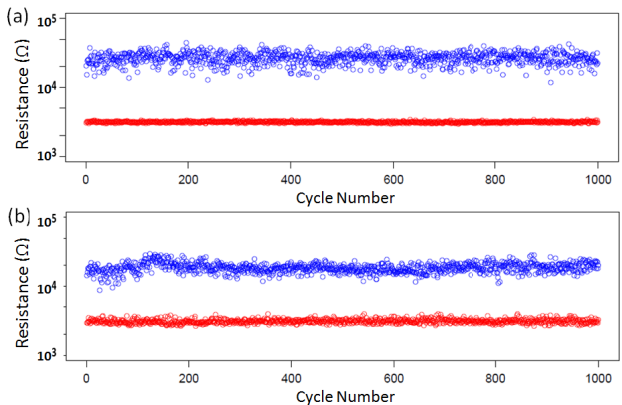

Fig. 10. HRS and LRS resistances measured in 1000 pulsed AC cycles in bench test. (a) Device 1 with Icomp = 350 µA. (b) Device 2 with Icomp = 500 µA.

The proposed method of BER analysis is also used in the off-line bench tests. Figure 10 shows the raw data of 1000 pulse AC write cycles measured by an off-line bench tester. Since the in-line and off-line testers were operated in different conditions, the RRAM write operation was stabilized differently; the result in Fig. 10 was measured with a set compliance current (Icomp) of 350 µA and 500 µA (set by an external transistor), compared to Icomp of 100 µA (set by Agilent 4073 parametric tester) used in the in-line result shown in Fig. 4. After using the logarithmic Q-Q fitting and formula (6), the measured BER curves for several devices are shown in Fig 11. This result demonstrates that RRAM BER curves may change with the write operation conditions, such as the pulse amplitudes, widths, and the compliance current values. Therefore understanding BER curves and their device to device variations at the specific circuit design points are essential for the device modeling and circuit simulations.

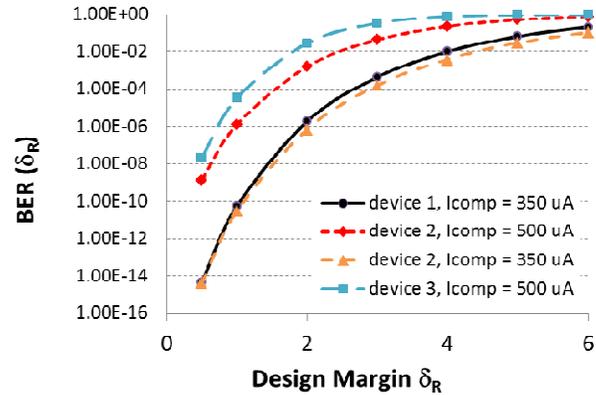

Fig. 11. BER curves of several device sites at various compliance currents.

The in-line-test BER characterization gives manufacturers a new capability for the early assessment and monitoring of the final product performances. Combined with programmable circuits or electrical fuse options, it can be used to set product error-correction-code (ECC) circuit configuration.

The proposed BER calculation can also be used in the build-in-self-test (BIST) for adaptive error correction and for the other memory products, including phase-change memory (PCM) and magnetic random access memory (MRAM).

## IV. CONCLUSION

HfO$_x$-based RRAM manufacturing variability and write BERs are investigated. A method is developed to characterize the design (sensing) margin dependent BERs at the device level using the in-line-test write cycle data. It enables early BER learnings of emerging memory technologies and will provide quantitative guidance to system and circuit designers.